\documentclass[a4paper,11pt]{article}
\usepackage{heppub}
\usepackage{amsthm}
\usepackage{amsmath}
\usepackage{graphicx}
\usepackage{subcaption}
\usepackage{slashed}
\usepackage{amssymb}
\usepackage{bbold}
\usepackage{float}
\usepackage{tikz}
\usepackage{cleveref}
\usetikzlibrary{arrows.meta,calc,decorations.markings,math,arrows.meta}
\usetikzlibrary{cd}
\usepackage{MnSymbol}
\newcommand{\QQ}{Q\overline{Q}}

\newcommand*\diff{\mathop{}\!\mathrm{d}}

\newcommand{\be}{\begin{eqnarray}}
\newcommand{\ee}{\end{eqnarray}}

\begin{document}
\title{Real-time Dynamics of the Schwinger Model as an Open Quantum System with Neural Density Operators}

\author[1,2]{Joshua Lin,}
\author[1,2,3]{Di Luo,}
\author[4]{Xiaojun Yao,}
\author[1,2]{and Phiala E. Shanahan}

\affiliation[1]{Center for Theoretical Physics, Massachusetts Institute of Technology, Cambridge, MA 02139, USA}
\affiliation[2]{The NSF AI Institute for Artificial Intelligence and Fundamental Interactions}
\affiliation[3]{Department of Physics, Harvard University, Cambridge, MA 02138, USA}
\affiliation[4]{InQubator for Quantum Simulation, Department of Physics, University of Washington, Seattle, WA 98195, USA}

\emailAdd{joshlin@mit.edu}
\emailAdd{diluo@mit.edu}
\emailAdd{xjyao@uw.edu}
\emailAdd{pshana@mit.edu}

\preprint{
    \vbox{\hbox{MIT-CTP/5679, IQuS@UW-21-073}}}

\abstract{
Ab-initio simulations of multiple heavy quarks propagating in a Quark-Gluon Plasma are computationally difficult to perform due to the large dimension of the space of density matrices. 
This work develops machine learning algorithms to overcome this difficulty by approximating exact quantum states with neural network parametrisations, specifically Neural Density Operators. 
As a proof of principle demonstration in a QCD-like theory, the approach is applied to solve the Lindblad master equation in the $1+1$d lattice Schwinger Model as an open quantum system. Neural Density Operators enable the study of in-medium dynamics on large lattice volumes, where multiple-string interactions and their effects on string-breaking and recombination phenomena can be studied. 
Thermal properties of the system at equilibrium can also be probed with these methods by variationally constructing the steady state of the Lindblad master equation. 
Scaling of this approach with system size is studied, and numerical demonstrations on up to 32 spatial lattice sites and with up to 3 interacting strings are performed.
}

\maketitle
 
\section{Introduction}

Understanding the mechanism by which quarks confine and deconfine is a longstanding theoretical problem \cite{Greensite:2011zz}.
Part of the puzzle is understanding the properties of the high-temperature deconfined phase of nuclear matter known as the Quark Gluon Plasma (QGP). 
A fundamental observable signature of QGP formation \cite{Matsui:1986dk} is the suppression of heavy quarkonia production at heavy-ion colliders.
Quantitatively understanding the observed suppression requires ab-initio simulations of the heavy quark dynamics and their hadronisation in the QGP.
In the high-temperature limit, the open quantum system framework describes this evolution as a first-order Lindblad master equation that can be numerically studied \cite{Akamatsu:2014qsa,Yao_2021,Akamatsu:2020ypb,Brambilla:2016wgg,Brambilla:2017zei,Scheihing-Hitschfeld:2023tuz}. 
As such, the development of numerical methods for the study of open quantum systems is of particular importance in this context.

A fundamental obstacle impeding progress into simulations of Lindbladian dynamics is the large dimension of the space of density matrices, which grows exponentially with the number of particles for a non-relativistic system, and exponentially in the volume for a relativistic system. 
Several approaches to this problem are under investigation, including Master Equation Unraveling \cite{theory,NGisin_1992}, and tensor-network parametrisations \cite{Sulz:2023ybe,PhysRevX.11.021035}.
Master Equation Unraveling has been applied to study the in-medium dynamics of a single heavy quark $Q$ or single heavy quarkonium $\QQ$ \cite{Akamatsu:2018xim,Miura_2020,Sharma_2020,PhysRevLett.68.580,Brambilla:2020qwo} where the dimension of the Hilbert space for wavefunctions is still manageable and exact representations of the quantum states can be used. 
A promising alternative approach is to use the neural network based approach of Neural Density Operators (NDOs), which variationally parametrise quantum states in far fewer parameters than the dimension of the space. 
Recent advances in computation and algorithms have allowed neural network methods to efficiently extract spectral data and Hamiltonian dynamics from a variety of physical systems \cite{Carleo_2017,Carleo:2019ptp}. Investigations of neural network methods for the more difficult problem of simulating open quantum systems, however, are still quite nascent, and early investigations have largely focused on open quantum spin-systems with local or close to local interactions \cite{Nagy_2019,Hartmann:2019rht,Vicentini_2019,Yoshioka_2019,Torlai_2018,Reh:2021wgn,Dugan:2023loz,luo2021autoregressive,Eeltink:2023jvi,Vicentini:2022jxq,Mellak:2022qzw}. 

This work studies the applicability of neural network methods, in particular NDOs, for investigating the properties and real-time dynamics of quantum field theories as open quantum systems. As a proof-of-principle demonstration, our study focuses on the $1+1$d lattice Schwinger Model~\cite{Schwinger:1962tp}, which shares many common features with QCD such as confinement and chiral symmetry breaking, and is often used as a testbed for new methods. When open boundary conditions are specified for the lattice theory, the gauge links can be integrated out, yielding a Hamiltonian with long-range spin interactions.
The unitary Hamiltonian dynamics of the Schwinger Model have been studied extensively with digital quantum circuits, tensor-network methods, and neural network methods \cite{Belyansky:2023rgh,Halimeh:2022pkw,Luo:2021ccm,Klco:2018kyo,Magnifico:2019kyj,Muschik:2016tws,Pichler:2015yqa,Zhang:2023hzr,Hauke:2013jga,Banuls:2016gid,Buyens:2016ecr,Buyens:2016hhu,Banuls:2016lkq,Zache:2018cqq,Zache:2021ggw,Farrell:2023fgd,Farrell:2024fit,Florio:2023dke,Angelides:2023bme,Schuster:2023klj,Florio:2023mzk}.
In this work, the NDO parametrisation \cite{Torlai_2018} is used to variationally parametrise density matrices. This allows numerical simulations of non-unitary dynamics for larger system sizes than would be possible with exact representations of the density matrices. The non-equilibrium dynamics of multiple interacting $\QQ$ pairs is simulated with the time-dependent Variational Monte-Carlo algorithm (tVMC) \cite{Hartmann:2019rht}. 
Algorithms to bootstrap NDOs from smaller lattice sizes to larger lattice sizes are also developed, which enables the non-equilibrium dynamics of the system to be probed on large lattice volumes. This paper is organized as follows: The Schwinger Model as an open quantum system on a lattice is reviewed in Section~\ref{sec:schwinger}. The NDO construction and tVMC algorithm are reviewed in Section~\ref{sec:nn_algo}. Results for Lindbladian dynamics simulated with NDO+tVMC are presented in Section~\ref{sec:real}. Thermal properties of the Schwinger model are investigated by variationally approximating the steady state of the Lindblad equation in Section~\ref{sec:steady}. In principle these methods are extendable to QCD simulations, as discussed in \Cref{sec:outlook}.

\section{Schwinger Model as an Open Quantum System on the Lattice}
\label{sec:schwinger}

The Schwinger model is a $U(1)$-gauge theory with a massive relativistic fermion (sometimes called an electron) in ${1+1}$d that shares many of the features of QCD, such as confinement and chiral symmetry breaking \cite{PhysRev.128.2425}. The Schwinger model as an open quantum system (coupled to a thermal environment) has been studied as a model of heavy quarkonia in the QGP \cite{deJong:2021wsd,Lee:2023urk}. A prototypical construction is to consider a Schwinger fermion $\psi$ coupled to a scalar field $\phi$:
\begin{equation}\label{eq:sch}
\mathcal{L} = -\frac{1}{4} F_{\mu \nu}F^{\mu \nu} + \overline{\psi} (i \slashed{D} - m)\psi + g \phi \overline{\psi}\psi  + \mathcal{L}_\phi.  
\end{equation}
The scalars $\phi$ are the environment degrees of freedom, and are traced out in the Quantum Brownian Motion limit leaving a Lindblad equation describing the remaining Schwinger degrees of freedom. Note that the gauge links are not traced out in this derivation of the Lindblad equation, although in $1+1$d the gauge links can be integrated out explicitly with open boundary conditions. 

To numerically simulate the Schwinger Model, the theory is discretised onto a lattice with $L$ points indexed by $0 \leq i,j,\cdots \leq (L-1)$ and lattice spacing $a$. In a staggered discretisation for the Schwinger fermion, the lattice has $\frac{L}{2}$ physical sites, with electrons occupying the even sites and positrons occupying the odd sites. A general Lindbladian evolution for such a system coupled to a thermal reservoir at temperature $T$ is parametrised (up to $O(1/T^2)$ effects) by a spatial coupling matrix $D_{ij}$ and operators $O_i$:
\begin{equation}
\label{eq:Lindblad}
\mathcal{L} \rho(t) := \frac{\diff \rho(t)}{\diff t} = -i \big[H,  \rho(t) \big] + a^2 \sum_{i,j} \, D_{ij} 
\left( \widetilde{O}_i \rho(t) \widetilde{O}^\dagger_j 
- \frac{1}{2} \big\{\widetilde{O}^\dagger_j \widetilde{O}_i, \rho(t) \big\} \right),
\end{equation}
\begin{equation}
\widetilde{O}_i \equiv O_i - \frac{1}{4T} \big[H, O_i \big] , \quad  
\widetilde{O}^\dagger_i \equiv  O_i + \frac{1}{4T} \big[H, O_i \big].
\end{equation}
All super-operators such as the Lindbladian $\mathcal{L}$ that act on the space of operators are labelled in calligraphic font. The coupling matrix $D_{ij}$ physically encodes the length-scale of correlations in the environment (the Debye screening length). The thermal state $\rho_T = \frac{1}{Z} e^{-\frac{H}{T}}$ satisfies $\mathcal{L} \rho_T = O\left({1}/{T^2}\right)$ so that the steady state of the Lindbladian evolution is approximately thermal. For the specific case of the Schwinger Model in temporal gauge (where the timelike gauge potential is zero)
with open boundary condition and spatial gauge links integrated out~\cite{Chakraborty:2020uhf}\footnote{The Hamiltonian constructed in Weyl gauge shown in \Cref{eq:schwinger_ham_1} is also equivalent to the Hamiltonian constructed in axial gauge with the spatial gauge potential set to zero~\cite{Farrell:2022wyt}.}, the Hamiltonian is the sum $H = H_\mathrm{kin.} + H_{\mathrm{mass}} + H_{\mathrm{elec.}}$, where:
\begin{equation}\label{eq:schwinger_ham_1}
    H_\mathrm{kin.} \equiv  \frac{1}{2a} \sum_{i=0}^{L-2} \big( \sigma_i^+ \sigma_{i+1}^- + \sigma_{i+1}^+ \sigma_{i}^- \big), \ \ H_\mathrm{mass}  \equiv m\sum_{i = 0}^{L - 1} O_i, \ \ H_\mathrm{elec.} \equiv \frac{a}{2} \sum_{i = 0}^L E_i^2,
\end{equation}
\begin{equation}\label{eq:schwinger_ham_2}
    O_i \equiv (-1)^i \frac{\sigma^z_i+1}{2a} , \quad E_i \equiv e \left(\ell_0 + \frac{1}{2} \sum_{j = 0}^{i - 1} (\sigma^z_j + (-1)^j) \right),
\end{equation}
where $m,e$ are the bare mass and electric charge respectively, and $\ell_0$ denotes the boundary condition, labelling the incoming electric flux at the $i=0$ lattice site. The fermionic degrees of freedom have been converted to spin-$\frac{1}{2}$ degrees of freedom with the Jordan-Wigner mapping \cite{Jordan:1928wi}. In this study, the boundary condition for the electric fields $E_i$ is chosen to be $\ell_0=0$.

The $D_{ij}$ matrix appearing in \Cref{eq:Lindblad} encodes the coupling between the Schwinger Model degrees of freedom and the environment degrees of freedom (in the example Lagrangian shown in \Cref{eq:sch}, this is a Yukawa coupling to an environment scalar). Its specific value will depend on the strength of the Yukawa coupling as well as the Lagrangian describing the environment degrees of freedom $\mathcal{L}_\phi$, and can in principle be computed perturbatively. 
As discussed in further detail in \Cref{app:qn}, $CP$-conservation of the total Hamiltonian shown in \Cref{eq:sch} places constraints on the form of the $D_{ij}$ matrix, in particular the Lindbladian $\mathcal{L}$ must be weakly $CP$-preserving ($[CP \otimes CP,\mathcal{L}] = 0$). 
Simulation results presented in \Cref{sec:res} will focus on two special limits:
\begin{equation}\label{eq:couplings}
\text{Delta Coupling}: D_{ij} = {D_0 \delta_{ij}}, \quad \text{Constant Coupling}: D_{ij} =  D_0 \mathbf{1}_{ij},
\end{equation}
where the delta coupling is point-like, and $\mathbf{1}_{ij}$ evaluates to $1$ for any choice of $\{i,j\}$ in order that the constant coupling has effectively infinite range.
The delta coupling is `weakly $CP$-preserving' ($[CP \otimes CP, \mathcal{L}] = 0$) whereas the constant coupling is `strongly $CP$-preserving' ($\mathcal{L}^\dagger(CP) = 0$) which leads to qualitatively different Lindbladian evolution behaviours.

\section{Neural Density Operators and Variational State Algorithms} 
\label{sec:nn_algo}

In this section, the neural network approach to simulating Lindbladian dynamics using NDOs is reviewed. There are several components of the approach:
\begin{itemize}
\item[-] The NDO construction, which is a variational parametrisation of density matrices $\rho : \mathbb{C}^N \to \mathcal{H} \otimes \mathcal{H}^*$, is reviewed in \Cref{subsec:NDO};
\item[-] In order to simulate Lindbladian evolution, the Time-Dependent Variational Principle is applied to vary the parameters of the NDO in order to best approximate the true evolution. The practical application of this approach is described in \Cref{subsec:tvmc};
\item[-] An approach to improve the efficiency of the initialization of the parameters of the NDO to prepare initial states is described in \Cref{subsec:init}. A transfer-learning procedure to transfer NDO-states prepared on a lattice of size $L$ to a lattice of size $2L$ is described. 
\end{itemize}

\subsection{Neural Density Operators}\label{subsec:NDO}

A large class of lattice discretisations of field theories (including the discretisation of the Schwinger Model used in this study) have a total Hilbert space that can be written as a tensor product over local Hilbert spaces $\mathcal{H} = \otimes_i \mathcal{H}_i$, where $i$ ranges over the lattice sites. The local Hilbert spaces $\mathcal{H}_i$ encode the local degrees of freedom (both fermionic and gauge degrees of freedom) that live at site $i$. In such a discretisation, the dimension of the total Hilbert space scales exponentially with the lattice volume $\mathrm{dim}(\mathcal{H}) = \Lambda^V$, where $\Lambda = \mathrm{dim}(\mathcal{H}_i)$ is the site dimension. As a result, it is prohibitively expensive to represent quantum states exactly on classical computers for all but small volumes. The problem is exacerbated when simulating open quantum systems, as density matrices naturally live in the space $\mathcal{H} \otimes \mathcal{H}^*$ which has dimension $\Lambda^{2V}$. This often limits the types of interactions that are practically simulatable: for example in the Open Schwinger Model the number and configuration of strings that can be simulated are limited by the volume of the lattice. 

\begin{figure}[t]
\includegraphics[scale=1]{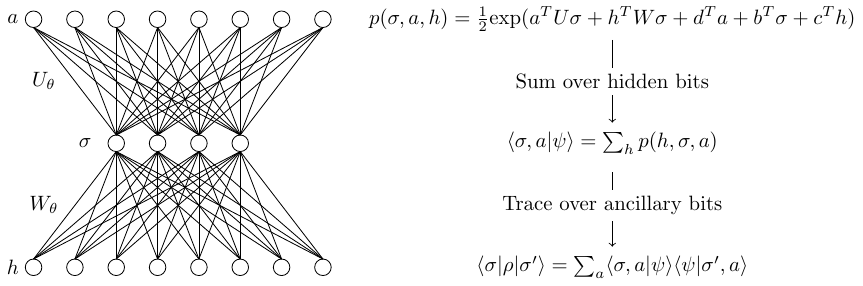}
\caption{Diagrammatic representation of the Neural Density Operator as presented in Refs.~\cite{PhysRevLett.120.240503,Hartmann:2019rht}. In the center are the physical spins $\sigma$, which take values $\sigma_i \in \{-1,+1\}$ for $i = 0,\dots,L-1$. Note that $\sigma$ here denotes the basis states, not Pauli matrices. There are also hidden bits $h_j \in \{-1,1\}$ and ancillary spins $a_k \in \{-1,1\}$ for $j = 0,\dots,M_h L -1$, $k = 0,\dots,M_a L-1$.}
\label{fig:NN}
\end{figure}

Neural network parametrisations of quantum states allow states in large-dimensional spaces to be represented compactly by far fewer parameters. 
The NDO shown in \Cref{fig:NN} is one particular neural network parametrisation  of density matrices $\rho : \mathbb{C}^N \to \mathcal{H} \otimes \mathcal{H}^*$ where $\mathbb{C}^N$ is the parameter-space \cite{PhysRevLett.120.240503,Hartmann:2019rht}. The parametrisation can be explicitly written as:
\begin{align}\label{eq:explicit_NN}
\log (\langle \sigma | \rho | \sigma' \rangle) &=  \sum_{j = 0}^{L-1} (b_j \sigma_j + b_j^* \sigma'_j ) + \sum_{k = 0}^{M_h L - 1} \mathcal{X}_k + \sum_{p=0}^{M_a L - 1} \mathcal{Y}_p, \\
\mathcal{X}_k &= \log \mathrm{cosh} \left( c_k + \sum_{j = 0}^{L-1} W_{kj} \sigma_j \right) + \log \mathrm{cosh} \left( c_k^* + \sum_{j = 0}^{L-1} W_{kj}^* \sigma'_j \right), \\
\mathcal{Y}_p &= \log \mathrm{cosh} \left( d_p + d_p^* + \sum_{j = 0}^{L-1} (U_{pj} \sigma_j + U_{pj}^* \sigma'_j ) \right). 
\end{align}
There are two positive hyperparameters $M_h, M_a$ (chosen such that $M_h L$ and $M_a L$ are integers) that control the size of the complex parameter space $\alpha$: 
\begin{equation}
\alpha = \{    W \in \mathbb{C}^{M_h L \times L}, U \in \mathbb{C}^{M_a L \times L}, b \in \mathbb{C}^L, c \in \mathbb{C}^{M_h L}, d \in \mathbb{C}^{M_a L} \}.
\end{equation}
For fixed $M_h,M_a$, the total number of complex parameters in $\alpha$ grows quadratically as $L$ increases. By construction the density matrix represented by \Cref{eq:explicit_NN} is Hermitian and positive definite, as it is defined as a trace over ancillary qubits of a pure-state wavefunction. 

In practice, it is conventional to represent the density matrix numerically as $\log \langle \sigma | \rho | \sigma' \rangle$, rather than $\langle \sigma | \rho | \sigma' \rangle$, as it increases numerical stability when parametrising matrices whose entries range over several orders of magnitude \cite{netket3:2021}. Another practical concern is that the definition of $\alpha$ as a set of complex parameters can lead to numerical instabilities as those parameters are varied (e.g., in optimization/training), due to the branch point singularities of the $\log \cosh$ function; a possible regulator is given by replacing the $\log \cosh$ functions appearing in the definition of \Cref{eq:explicit_NN} with a regulated $\log \cosh$ function:
\begin{equation}
\text{RLC}(z) := \frac{\log \cosh (\frac{\tanh(|z|)}{|z|} z)}{
(\frac{\tanh(|z|)}{|z|})},
\end{equation}
which no longer has any branch-cuts, has the same behaviour near $z = 0$ as $\log \cosh$, but is no longer holomorphic. As a result, this modification causes certain values of the parameters $\alpha$ to no longer yield positive-definite density matrices, possibly introducing systematic uncertainties. 

\subsection{Time-Dependent Variational Monte Carlo}\label{subsec:tvmc}

A method to simulate the real-time dynamics of quantum states parametrised by some set of variational parameters $\alpha$ is the Time Dependent Variational Principle (TDVP) \cite{1981}. In the case of Lindbladian dynamics with $\frac{d}{dt} \rho(t) = \mathcal{L} \rho$, minimizing the $L^2$-norm (denoted $\| \cdot \|_2$) between the true time derivative $\mathcal{L} \rho$ and the difference induced by an  infinitesimal variation in the neural network parameters $\alpha$ gives:
\begin{align}\label{eq:tdvp}
&\min_{\dot{\alpha}} \left\| \sum_{i}  \dot{{\alpha}}_i  \frac{\partial \rho_a}{\partial \alpha_i} - \sum_b \mathcal{L}_{ab} {\rho_b}  \right\|_2 \  \implies \  \dot{{\alpha}} = S^{-1} f, \\
(\mathcal{O}_i)_{ab} := \delta_{ab}\frac{\partial \ln(\rho_a)}{\partial \alpha_i}&, \quad S_{ij} := \rho^\dagger \mathcal{O}_i^\dagger \mathcal{O}_j \rho + \rho^\dagger \mathcal{O}_j^\dagger \mathcal{O}_i \rho, \quad f_i := \rho^\dagger \mathcal{O}_i^\dagger \mathcal{L} \rho + \rho^\dagger \mathcal{L}^\dagger \mathcal{O}_i \rho,
\end{align}
where $\rho$ denotes the vectorised density matrix, so that the Lindbladian super-operator $\mathcal{L}$ acts on it as a regular matrix $\rho \mapsto \mathcal{L} \rho$. 
$S$ is the Quantum Geometric Tensor (QGT) \cite{Berry:1989ai}, which is the pullback of the $L^2$ metric $\| \cdot \|_2$ on the vectorized space of density matrices, along the variational parametrisation map $\rho : \mathbb{C}^N \to \mathcal{H} \otimes \mathcal{H}^*$. The QGT parametrises the change of the physical density matrix under an infinitesimal change in the parameters of the neural network. 

In any variational parametrisation of a large Hilbert space, the forces $f_i$ and the QGT $S_{ij}$ are expensive to compute exactly. Instead, it is possible to construct Monte-Carlo Markov Chains (MCMCs) to estimate these quantities. The combination of the TDVP algorithm with stochastic estimation is known as the time-dependent Variational Monte Carlo (tVMC). 
The forces $f_i$ and QGT $S_{ij}$ from \Cref{eq:tdvp} can be rewritten as:
\begin{equation}\label{eq:SF}
S_{ij} = Z \sum_a \frac{\|\rho_a\|^2}{Z} 2 \mathrm{Re}\left( (\mathcal{O}_i^\dagger)_{aa} (\mathcal{O}_j)_{aa} \right), \quad f_i = Z \sum_{a} \frac{\|\rho_a\|^2}{Z} 2 \mathrm{Re}\left(( \mathcal{O}_i^\dagger)_{aa} \sum_b \frac{\mathcal{L}_{ab} \rho_b}{\rho_a} \right), 
\end{equation}
where $Z = \sum_a \|\rho_a\|^2$ is a normalization constant. When evaluating the time derivative of the parameters $\dot{\alpha} = S^{-1} f$, the factors of $Z$ cancel, and \Cref{eq:SF} can be stochastically estimated using a MCMC that draws samples from the basis states of the doubled Hilbert space according to the probability distribution $\|\rho_a^2\|/Z$. To regulate the possible divergences that may appear in inverting $S$, a small diagonal offset is added ($S_{ij} \mapsto S_{ij} + \epsilon \delta_{ij}$) which ensures its invertability. Note that this offset should be taken to be as small as possible, as it introduces a systematic error into the simulated Lindbladian evolution. Physical observables of interest $O$ (such as the chiral condensate or electric field for the Open Schwinger Model) can be stochastically estimated via the following identity (which is expressed in terms of basis elements of the original, un-doubled Hilbert space):
\begin{equation}
\langle O \rangle = \frac{\mathrm{Tr}(\rho O)}{\mathrm{Tr}(\rho)} = \sum_\sigma \left( \frac{\langle \sigma | \rho | \sigma \rangle}{\mathrm{Tr}(\rho)} \right) \left( \frac{\langle \sigma | \rho O | \sigma  \rangle}{\langle \sigma | \rho | \sigma \rangle}\right), 
\end{equation}
which requires constructing a separate MCMC to sample from the diagonal of the density matrix according to the probability distribution $|\sigma\rangle \sim \langle \sigma | \rho | \sigma \rangle$. 

Regardless of the initial state, in simulating the Lindbladian dynamics as $t \to \infty$ the system is driven towards its steady-state. This algorithm is also known as the Stochastic Reconfiguration algorithm, and can be recast as a preconditioned version of regular gradient descent \cite{Sorella_1998,Sorella_2007}. Adding larger diagonal shifts to $S$ in the Stochastic Reconfiguration framework brings the updates closer to regular gradient descent, and does not bias the machine-learned steady state (in contrast to the systematic uncertainty that it introduces in the Lindbladian evolution itself). For finite values of $M_h,M_a$, the NDO parametrisation will not capture the full space of density matrices, which introduces systematic errors in both the estimated Lindbladian evolution, as well as the estimated steady states. Furthermore, the Monte-Carlo estimations of the QGT and forces come with inherent statistical uncertainty, that only vanishes in the limit that the number of samples used is taken to infinity. 

\subsection{Initialization}\label{subsec:init}

Studying Lindbladian dynamics first requires preparation of initial states of interest. It is possible, for example, to explicitly set the weights of the NDO to prepare given product states. Consider the case of preparing a product state $\rho_{\vec{a}} = |a_0 a_1 \cdots a_{L-1} \rangle \langle a_0 a_1\cdots a_{L-1} |  $ (where $\mathrm{Tr}(\sigma_i^z \rho_{\vec{a}}) = a_i$). By setting:
\begin{equation}
W_{kj} = \lambda a_j \delta_{k0} + O(\epsilon), \quad c_k = \lambda L \delta_{k0} + O (\epsilon),
\end{equation}
for some choice of $\lambda$, with all other parameters initialized to $O(\epsilon)$ for some $\epsilon \ll 1$, the weights $W_{kj}$ will pick out the $\rho_{\vec{a}}$-state, and the bias-term $c_k$ will break the symmetry between $\rho_{\vec{a}}$ and $\rho_{-\vec{a}}$, so that the NDO represents the state $\rho_\mathrm{NDO} = \rho_{\vec{a}} + O(e^{-\lambda})$.

\begin{figure}[t]
\begin{center}
\includegraphics{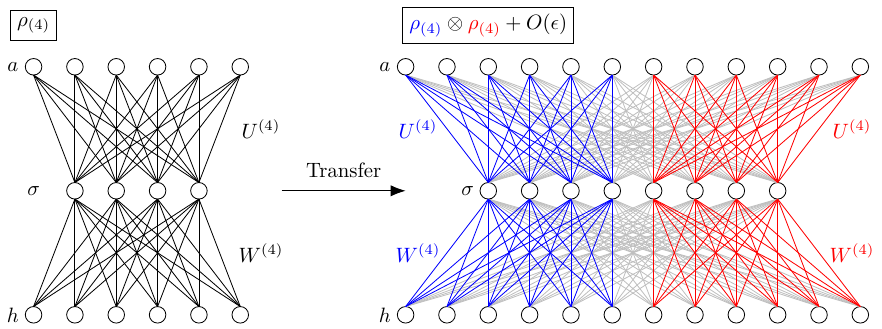}
\be 
\begin{aligned}
&\qquad\qquad\ \  U^{(8)} = \begin{bmatrix} U^{(4)} & 0 \\ 0 & U^{(4)} \end{bmatrix} + O(\epsilon), \quad   W^{(8)} = \begin{bmatrix} W^{(4)} & 0 \\ 0 & W^{(4)} \end{bmatrix} + O(\epsilon), \\ &b^{(8)} = \begin{bmatrix} b^{(4)} & b^{(4)} \end{bmatrix} + O(\epsilon) , \quad c^{(8)} = \begin{bmatrix} c^{(4)} & c^{(4)} \end{bmatrix} + O(\epsilon) , \quad d^{(8)} = \begin{bmatrix} d^{(4)} & d^{(4)} \end{bmatrix} + O(\epsilon) 
\nonumber
\end{aligned}
\ee 
\end{center}
\caption{The procedure for transferring weights from a network optimized to represent a density matrix defined on $L$ lattice sites to a network representing one on $2L$ sites (for any fixed values of $M_a,M_h$). Shown in the figure is a schematic for transferring weights $\alpha^{(4)} = \{U^{(4)},W^{(4)},b^{(4)},c^{(4)},d^{(4)}\}$ from an NDO with  $M_a = M_h = 1.5$ on an $L = 4$ lattice to weights $\alpha^{(8)}$ for an $L = 8$ NDO, with the weight definitions from \Cref{eq:explicit_NN}. The $\epsilon \ll 1$ are small random numbers. }
\label{fig:transfer}
\end{figure}

When probing the steady-state properties of a system with the Stochastic Reconfiguration algorithm, initializing the NDO to an arbitrary state will lead to the correct steady state. 
However, initializing the model to a state close to the true solution can reduce simulation times drastically. 
One strategy is to exploit the approximate translational symmetry of the system, which is broken by $O(e^{-L \Delta E})$ effects due to the open boundary conditions, where $\Delta E$ is the gap between the ground state and the first excited state.
The approximate translational symmetry implies that the steady state on a lattice of size $2L$ has approximately the same structure on the left and right halves of the lattice as the steady state on a lattice of size $L$ with the same bare parameters. 
This motivates a transfer learning procedure where an NDO on a size $L$ lattice is first optimized to the steady state, with the resulting weights then used to initialize an NDO on a size $2L$ lattice, as shown in \Cref{fig:transfer}. 
By stitching together two copies of the smaller NDO together, the structure of the steady state is preserved on both halves, and the Stochastic Reconfiguration algorithm need only learn the coupling between the halves.

\section{Lindbladian Dynamics of Strings and Steady State Properties}
\label{sec:res}

NDOs as reviewed in \Cref{sec:nn_algo} are used in this section to variationally parametrise density matrices of the Schwinger Model as an open quantum system: 
\begin{itemize}
\item[-] To investigate the Lindbladian dynamics of strings in \Cref{sec:real}, product states containing string configurations are prepared and Lindbladian evolution is performed using tVMC. Comparisons to exact results are shown on small lattice sizes, and the scaling of the method as lattice size is increased is studied. Finally, Lindbladian evolution of string states on an $L = 20$ lattice are computed with tVMC, for which exact diagonalization methods are infeasible.
\item[-] The thermal behaviour of the system is studied in \Cref{sec:steady} by extracting the steady state of the Lindblad equation. The learned state is compared to exact diagonalization results on an $L = 4$ lattice, before being bootstrapped up to an $L = 32$ volume lattice (see \Cref{fig:transfer}). The key observable is the chiral condensate, which measures the chiral symmetry breaking and is the prototypical order parameter for the confinement/deconfinement phase transition. By performing a scan in the physical parameters, it is confirmed that the learned NDO state is sensitive to this phase transition. 
\end{itemize}

\subsection{Lindbladian Dynamics}
\label{sec:real}

\begin{table}[t]
\begin{center}
\begin{tabular}{l|l|l|l|l|l|}
\cline{2-6}
   & $a$   & $m$   & $e$   & $T$   & $D_{ij}$  \\ \hline
\multicolumn{1}{|l|}{Parameter set $1$ (delta):}            & $1$   & $0$   & $0.5$ & $10$  & $0.15 \ \delta_{ij}$ \\ \hline
\multicolumn{1}{|l|}{Parameter set $2$ (delta):}    & $0.6$ & $0.5$ & $2$   & $1.5$ & $0.72 \ \delta_{ij}$    \\ \hline
\multicolumn{1}{|l|}{Parameter set $2$ (constant):} & $0.6$ & $0.5$ & $2$   & $1.5$ & $0.216 \ \mathbf{1}_{ij}$              \\ \hline
\end{tabular}
\end{center}
\caption{Bare parameter sets for the Lindblad operator described in \Cref{sec:schwinger}, which are simulated at various lattice sizes with NDO+tVMC in \Cref{sec:real}.}
\label{tab:bare}
\end{table}

The Lindbladian dynamics of the Open Schwinger Model as presented in \Cref{sec:schwinger} is numerically studied with the bare parameters shown in \Cref{tab:bare}. Parameter set $1$ is in a regime where Lindbladian evolution of string states leads to string breaking, and matches the parameters studied in an earlier work \cite{Lee:2023urk}. 
Parameter set $2$ is tuned such that the Lindbladian dynamics starting with string product states of length $3$, and with the two different types of couplings (see \Cref{eq:couplings}), are approximately equal for small times but show deviations when simulated for long time intervals.
The initial states studied are product states describing $n$ electrons and $n$ positrons:
\begin{equation}
|e^+(x_1) \cdots e^+(x_n) e^-(y_1) \cdots e^- (y_n)\rangle = \sigma^-_{2x_1 + 1} \cdots \sigma^-_{2x_n + 1} \sigma^+_{2y_1} \cdots \sigma^+_{2y_n} |\downarrow \uparrow \cdots \downarrow \uparrow \rangle.
\end{equation}
Here, $x_i,y_i$ label physical sites rather than staggered sites, and the state $|\downarrow \uparrow \cdots \downarrow \uparrow \rangle$ is the free vacuum product state expressed in the $Z$-computational basis. Observables that are useful for visualising the real-time dynamics are the dimensionless chiral condensate and the electric field expressed in units of $e$: 
\begin{equation}
\overline{\psi} \psi := \frac{1}{L} \sum_{i=0}^{L-1} (-1)^i \frac{\sigma_i^z + 1}{2}, \quad E_i = \frac{1}{2}\sum_{j = 0}^{i-1} (\sigma_j^z + (-1)^j).
\end{equation} 
When initialising to product states containing string configurations, the time-evolved electric field $\langle E^\mathrm{string}_i \rangle(t)$ contains large vacuum fluctuations, which can be removed by subtracting the electric field of the time-evolved free vacuum state $|\downarrow \uparrow \cdots \downarrow \uparrow\rangle$, giving $\langle E^\mathrm{vac. sub.}_i \rangle(t) := \langle E^\mathrm{string}_i \rangle(t) - \langle E^\mathrm{vacuum}_i \rangle (t)$. For illustration purposes it is also useful to define a shifted electric field $\tilde{E}_i(t)$, which has been shifted by its value at $t = 0$, $\langle \tilde{E}_i \rangle (t) := \langle E_i \rangle(t) - \langle E_i \rangle(0)$.

\begin{figure}[t]
\hspace*{-0.2cm} 
\includegraphics[scale=1]{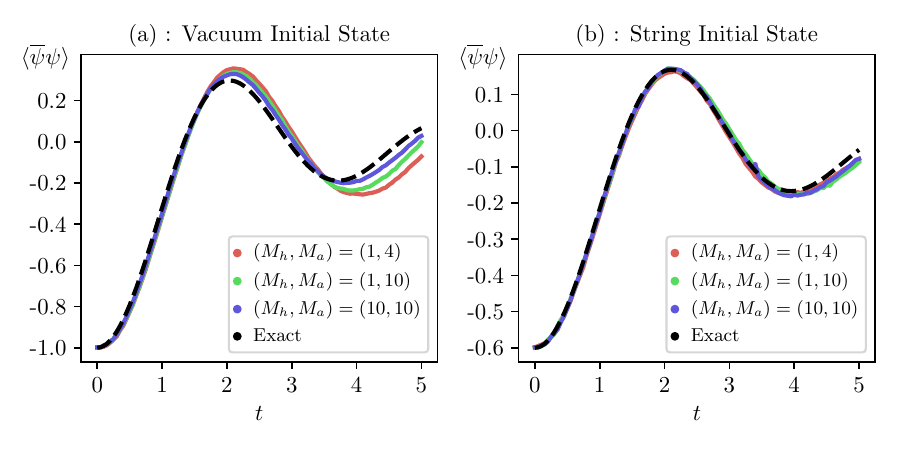}
\caption{Studies of parameter set $1$, on an $L = 10$ lattice. (a) The Lindbladian dynamics of the chiral condensate $\langle \overline{\psi} \psi\rangle$ computed with NDO's of various sizes, compared with the exact result. (b) The same, but with the string state $\sigma^-_{3} \sigma^+_6 | \downarrow \uparrow \cdots \downarrow \uparrow \rangle $ as the initial state. }
\label{fig:row1}
\end{figure}

\Cref{fig:row1} shows a comparison of $\langle \overline{\psi} \psi \rangle(t)$ between results computed with NDO-tVMC and results computed using exact representations, for parameter set 1 on an $L = 10$ lattice. Both NDO-tVMC and the exact results use the 4-th order Runge-Kutta integrator with $dt = 0.05$ to integrate the Lindblad equation. The diagonal shift regulator for the Quantum Geometric Tensor inversion is set to $5 \cdot 10^{-5}$. A total of $2^{17}$ samples were used in both MCMCs when estimating observables/forces, and the size of the NDO was varied as labelled in the figure. It was observed that vacuum fluctuations cause systematic errors in the computed Lindbladian evolutions of the chiral condensate that grow in time, which are relatively suppressed for string product states compared to vacuum product states. \Cref{fig:row1}(a) demonstrates that the network size significantly affects the systematic uncertainty in the tVMC-NDO algorithm, with larger network sizes resulting in smaller errors due to the higher representation power. 

\begin{figure}[t]
\begin{center}
\includegraphics[scale=1]{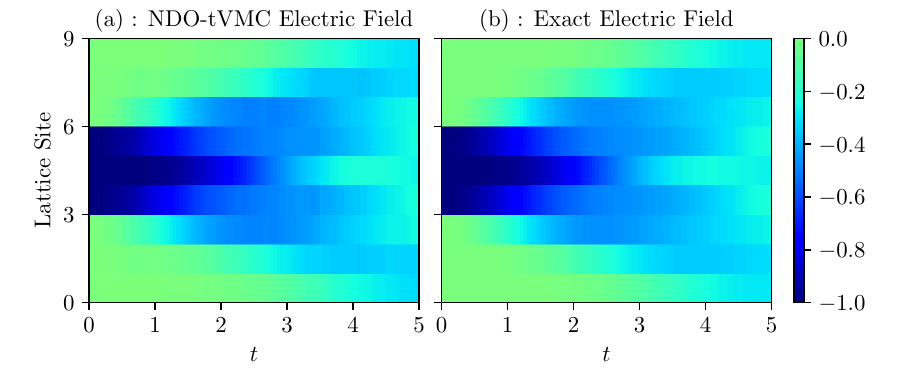}
\caption{The vacuum-subtracted electric fields as a function of time, computed for parameter set 1 with string state $\sigma^-_{3} \sigma^+_6 | \downarrow \uparrow \cdots \downarrow \uparrow\rangle $. (a) The result computed with a $(M_h,M_a) = (10,10)$ NDO which has been rolling-averaged over three timesteps to reduce statistical noise. (b) The exact result. }\label{fig:ecomp}
\end{center}
\end{figure}

\begin{figure}[t]
\begin{center}
\hspace*{0cm} 
\includegraphics[scale=1]{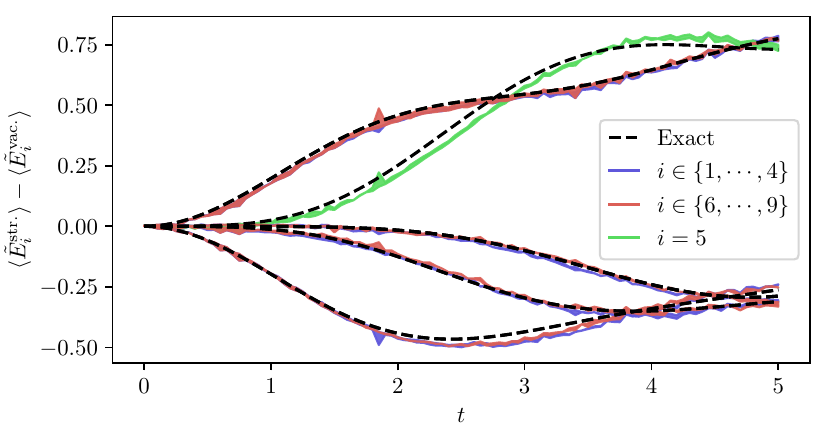}
\caption{The same data as \Cref{fig:ecomp}, with the shifted vacuum-subtracted electric field at each lattice site shown as a separate line. The NDO-tVMC electric fields measured on the left/right half of the lattice are denoted by blue/red bands. The green band corresponds to the electric field link exactly in the center of the lattice.}\label{fig:row3}
\end{center}
\end{figure}

\Cref{fig:ecomp} compares the vacuum subtracted electric fields $\langle E^\mathrm{string}\rangle(t) - \langle E^\mathrm{vacuum}\rangle(t)$ between the NDO-tVMC determination (using the largest network $(M_h,M_a) = (10,10)$) and the exact result. Qualitative agreement is observed between the solutions; for a more quantitative comparison, the shifted electric fields $\langle \tilde{E}_i^\mathrm{string}\rangle(t) - \langle \tilde{E}_i^\mathrm{vacuum} \rangle(t)$ are compared between the NDO-tVMC calculation and exact results in \Cref{fig:row3}. Due to the fact that the $\delta$-coupling is weakly $CP$-conserving, the electric fields satisfy the symmetry $\langle E_{i} \rangle(t) = \langle E_{L - i}\rangle(t)$. \Cref{fig:row3} demonstrates that the NDO-tVMC algorithm has preserved this symmetry well in the simulated Lindbladian dynamics, as the results of measurements of the electric field on each half of the lattice are consistent. 

\begin{figure}[t]
\begin{center}
\includegraphics[scale=1]{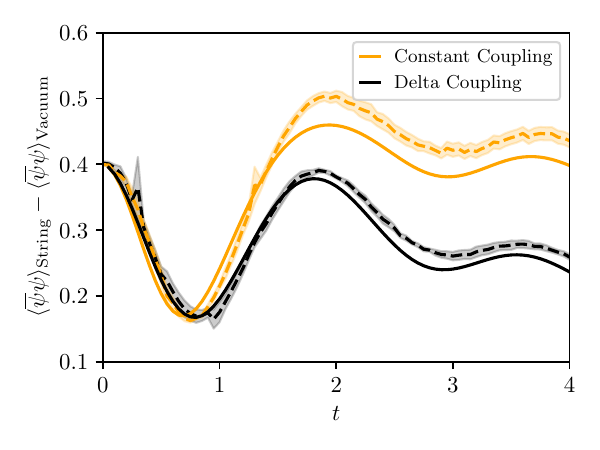}
\caption{Vacuum-subtracted chiral condensates computed with NDO-tVMC (dashed lines) for the two types of couplings in parameter set 2, with initial string state $\sigma^-_{3} \sigma^+_6 | \downarrow \uparrow \cdots \downarrow \uparrow\rangle $. Shaded bands indicate statistical uncertainty estimated by MCMC, added in quadrature with the uncertainty computed as the spread over three separate runs of the simulation. Exact results (solid lines) are shown for comparison. Computed on an $L = 10$ lattice, with $(M_h,M_a) = (10,10)$. }\label{fig:cd}
\end{center}
\end{figure}

The NDO-tVMC method is able to simulate different types of coupling matrices $D_{ij}$. Parameter set $2$ is specifically tuned such that the vacuum-subtracted chiral condensate has similar Lindbladian dynamics for small times when comparing the delta coupling to the constant coupling with the initial string state $\sigma_3^- \sigma_6^+ |\downarrow \uparrow \cdots \downarrow \uparrow\rangle$---\Cref{fig:cd} shows that they share the same first oscillation, but diverge in behaviour afterwards. As the constant coupling is strongly-CP conserving, there are two distinct steady states and the long-time dynamics of the constant coupling dynamics does not match the long-time dynamics of the delta coupling dynamics. \Cref{fig:cd} demonstrates the ability of the NDO-tVMC to distinguish between the two types of coupling.

\begin{figure}[t]
\begin{center}
\includegraphics[scale=1]{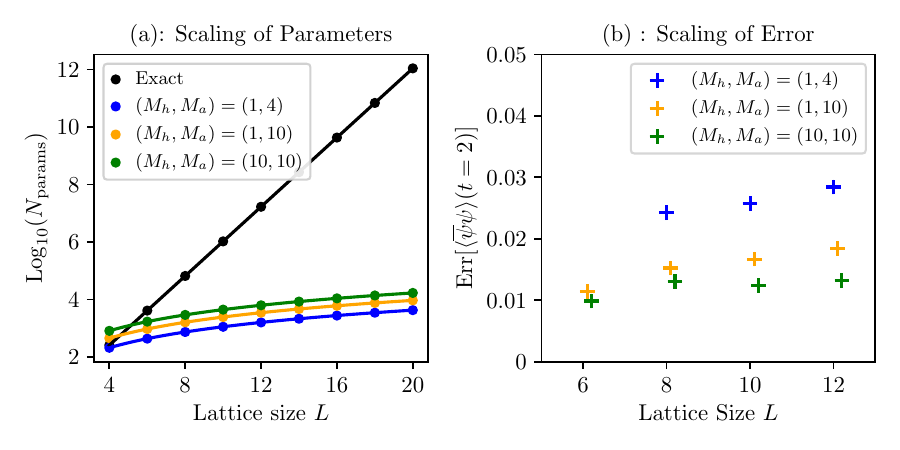}
\caption{(a) Scaling of the number of parameters in the NDO construction at fixed $(M_h,M_a)$ with the size of the lattice $L$, compared to the complex dimension of $\mathcal{H} \otimes \mathcal{H}^*$. The three trajectories shown are the same choices used in the comparison of \Cref{fig:row1}. (b) Scaling of errors at $t = 2$ for parameter set $1$ and the vacuum initial state as a function of the lattice size $L$. }\label{paramscale}
\end{center}
\end{figure}

\begin{figure}[t]
\begin{center}
\includegraphics[scale=1]{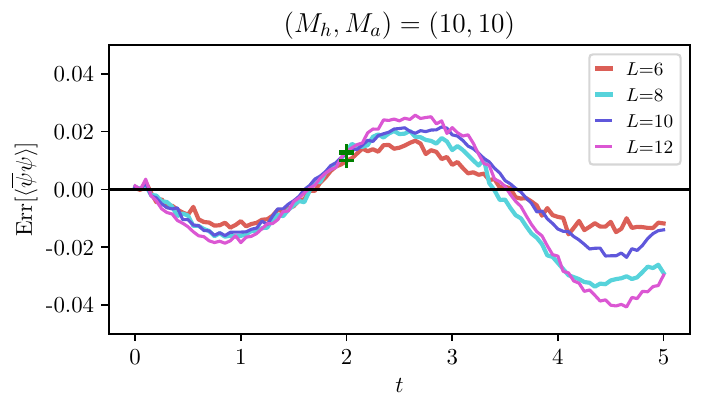}
\end{center}
\caption{For the largest neural network shapes tested ($(M_h,M_a) = (10,10)$), the normalised error in the chiral condensate for the vacuum initial state is shown as a function of time for various lattice sizes. The green markers denote the same points as those shown with the same symbol in \Cref{paramscale}(b). }\label{fig:rte}
\end{figure}

One of the advantages of the NDO approach is that it can be applied to lattice sizes $L$ where exact calculations are computationally intractable. However, it is important that the uncertainties arising from the truncated parameterization of the NDO can be controlled or estimated. In particular, for an NDO parameterisation with fixed $(M_h,M_a)$, the number of complex parameters in the model grows quadratically with the lattice size, whereas the dimension of the space of density matrices grows exponentially, as shown in \Cref{paramscale}(a). The effects of this (increasingly severe) truncation can be quantified in the regime of lattice sizes $L$ where exact calculations are possible, in order to understand the scaling of systematic effects.
\Cref{paramscale}(b) shows the scaling of the normalised error in the chiral condensate for the vacuum initial state evaluated at $t = 2$:
\begin{equation}
\mathrm{Err}[\langle \overline{\psi} \psi \rangle (t = 2)] := \frac{1}{2} (\langle \overline{\psi} \psi \rangle_{\mathrm{NDO}}(t = 2) - \langle \overline{\psi} \psi \rangle_{\mathrm{exact}}(t=2) )
\end{equation}
as a function of the lattice size $L$ holding all other hyperparameters constant (number of samples used in the MCMCs, regulator choice for the QGT inversion). The factor of $\frac{1}{2}$ accounts for the normalisation as $\langle \overline{\psi} \psi \rangle \in [-1,1]$. The time $t = 2$ is chosen as the peak of the first oscillation in the time-evolved chiral condensate, as observed in \Cref{fig:row1}(a). As shown in \Cref{paramscale}(b), at fixed time $t = 2$ and for fixed $(M_h,M_a)$ the scaling of the error as a function of lattice size is fairly mild, suggesting that the accuracy of the NDO-tVMC algorithm can be controlled as the lattice size is increased to larger values where exact results are not available. 
Other sources of systematic error in the NDO-tVMC algorithm include the number of samples used in the MCMC chains \cite{Sinibaldi:2023yoq}, choice of regulator for the Quantum Geometric Tensor inversion \cite{Schmitt_2020}, and the timestep used in ODE integrator.  

The normalised error in the chiral condensate for the vacuum initial state is shown as a function of simulation time in \Cref{fig:rte}, for a variety of lattice sizes with the largest tested NDOs, $(M_h,M_a) = (10,10)$. The approximate curve collapse for times $t \leq 2$ once again suggests mild scaling of errors as the lattice-size is increased at early to intermediate times during the evolution of the system. At later simulation times, the scaling of errors as a function of lattice size is less controlled; note, however, that there is no simple relationship between the size of the error for different lattice sizes. Although the time $t = 2$ corresponds to a physically relevant timescale (the first peak of the chiral condensate evolution for the vacuum initial state), it does not correspond to a maximum/minimum in $\mathrm{Err}[\langle \overline{\psi} \psi \rangle]$. 

\begin{figure}[t]
\centering
\hspace*{-0.5cm}
\includegraphics[scale=1]{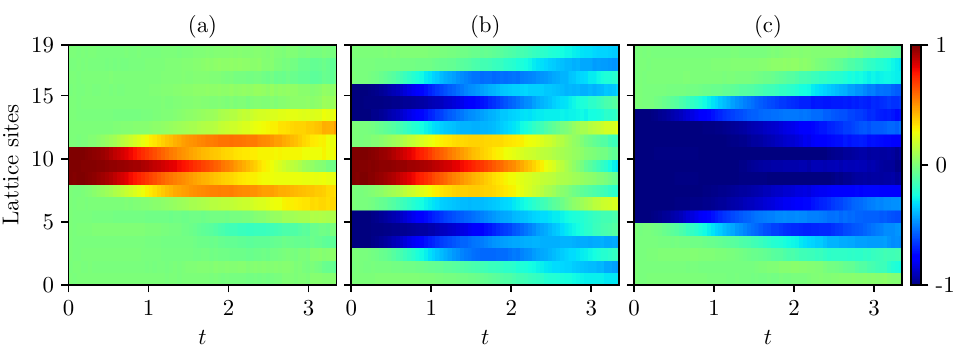}
  \caption{The vacuum-subtracted electric field as a function of time for various initial states, for parameter set $1$ on an $L = 20$ lattice. (a) Initial state is a single short string $\sigma^+_8 \sigma^-_{11} |\downarrow \uparrow \cdots \downarrow \uparrow \rangle$. (b) Initial state is three interacting strings $\sigma^-_3 \sigma^+_6 \sigma^+_8 \sigma^-_{11} \sigma^-_{13} \sigma^+_{15} |\downarrow \uparrow \cdots \downarrow \uparrow \rangle$. (c) Initial state is a single long string $\sigma^-_5 \sigma^+_{14} | \downarrow \uparrow \cdots \downarrow \uparrow \rangle$. }
  \label{fig:20}
\end{figure}

To demonstrate the utility of the NDO-tVMC approach for larger lattice sizes where exact results are computationally intractable to achieve, the Lindbladian dynamics of various initial states are simulated on an $L = 20$ lattice, as shown in \Cref{fig:20}.
The real dimension of the space of density matrices is approximately $3 \cdot 10^{10}$, which was parametrised by $9280$ real parameters corresponding to $M_h = 1, M_a = 4$ in the NDO construction. In \Cref{fig:20}(a) an initial $e^+ e^-$ pair joined by a short string splits into two strings due to the pair creation of an $e^+ e^-$ pair in-between the two fermions, which is the same physical scenario simulated in \Cref{fig:ecomp}, but on a larger lattice size where finite volume effects are suppressed. \Cref{fig:20}(b) illustrates that the dynamics of multi-string states can be investigated, at the same computational cost as simulating the dynamics of single-string states. Finally, \Cref{fig:20}(c) shows the dynamics of a long string, which can not be investigated on smaller lattice sizes. 

\subsection{Steady State Properties}
\label{sec:steady}

\begin{figure}[t]
\hspace*{-0.5cm}
\includegraphics[scale=1]{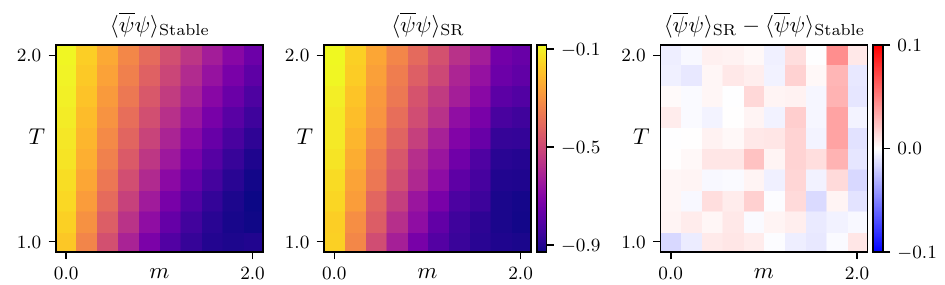}
\caption{Comparison between the exact chiral condensate of the steady state (left), with the chiral condensate computed by the Stochastic Reconfiguration algorithm (center) as a function of temperature $T$ and mass $m$. The unnormalised difference between the two is shown in the right figure.}
\label{fig:4scan}
\end{figure}

Extracting the steady state solution $\rho_\mathrm{stab.}$, which by definition satisfies:
\be 
\mathcal{L} \rho_\mathrm{stab.} = 0 \quad \implies \quad  \rho_\mathrm{stab.} = \mathrm{exp}(-H/T) + O(T^{-2}),
\ee 
allows thermal properties of the system to be probed. As discussed in \Cref{sec:nn_algo}, Stochastic Reconfiguration with a finite diagonal shift as a regulator for the Quantum Geometric Tensor inversion is used to drive the NDO towards the steady state. Transfer-learning allows weights on smaller lattice sizes to be used as initialization for training on larger lattice sizes, which both reduces training costs and increases the reliability of results, as described in \Cref{fig:transfer}. For a small $L = 4$ lattice size, exact results for the steady state are easily computable; comparisons of the chiral condensate as determined with Stochastic Reconfiguration to the exact results are shown in \Cref{fig:4scan}. Results shown are computed for a compact parameterization with $(M_h,M_a) = (1,1)$. A diagonal shift of $0.01$ is used, with 16000 samples distributed over 16 parallel chains for both the MCMC chains, and the system is trained for 6000 steps. The Lindbladian used is the delta-coupling Lindbladian $D_{ij} = \delta_{ij}$, which is chosen to ensure a unique steady state solution (as opposed to the constant coupling which has two separate steady states). The chiral symmetry breaking transition can be seen by the nonzero chiral condensate as the temperature is decreased, or as the bare mass is increased (explicit chiral symmetry breaking). 

\begin{figure}[t]
\begin{center}
\hspace*{-1.5cm}
\includegraphics[scale=1]{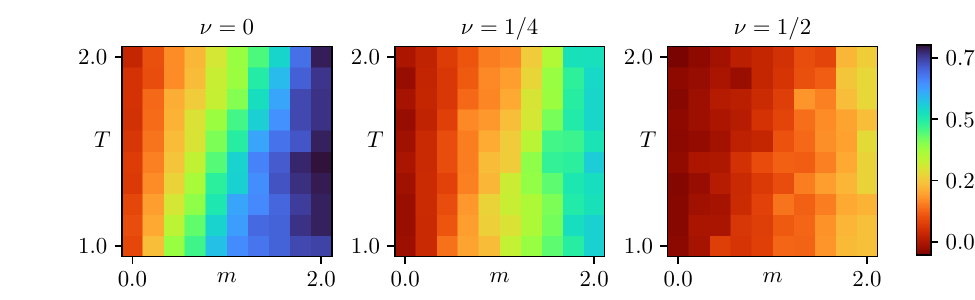}
\end{center}
\caption{Measurement of the chiral condensate of the learned steady state solution on a $L = 32$ lattice size. The three different columns show the system at different filling fractions $\nu = \frac{N}{L}$.  }
\label{fig:qq}
\end{figure}

The steady state of the $L=4$ system is used to bootstrap to results on larger lattice sizes where exact results can not be achieved; the state is transferred to an $L = 8$ lattice size, and retrained. This process is repeated twice, yielding an approximation of the steady state on an $L = 32$ lattice size. In the transfer learning, 8000 samples are distributed over 128 parallel chains, and gradients are clipped to have maximum $L^2$ norm $1$ \cite{pascanu}. The Stochastic Reconfiguration regulator parameter is set to $\epsilon = 0.1$ for all the training sets. \Cref{fig:qq} shows the behaviour of the measured chiral condensate as a function of $T,m$, and the filling fraction $\nu := \frac{N}{L}$ where $N = \sum_{i = 0}^{L-1} \sigma_i^z$ is the net particle number. The filling fraction $\nu$ varies in the range $\nu \in [-1,1]$. By $CP$-symmetry, the measured chiral condensate is symmetric about $\nu = \frac{1}{2}$ when all other bare parameters are held fixed. 

\section{Conclusions and Outlook}
\label{sec:outlook}

Neural network quantum state methods offer a promising approach to lowering the cost of classical simulations of open quantum systems. 
This work demonstrates the application of the NDO ansatz to the $1+1$d Schwinger Model as an open quantum system. 
Combined with tVMC, Lindbladian dynamics was simulated on various lattice sizes and with different bare Lindbladian parameters.
By comparing machine-learned results to exact results on small lattice sizes, the systematic errors introduced by the NDO-tVMC method were found to scale mildly as the lattice size is increased, for fixed neural network shape. 
This allows for systematically improvable simulation of Lindbladian dynamics at larger lattice sizes than possible with exact representations of the density matrices. 
As a demonstration, Lindbladian evolution of various string states was simulated on an $L = 20$ lattice, which allows for investigation of physical scenarios that are not possible to simulate on smaller lattice sizes, such as the interaction between multiple strings and fermion pairs. 
By simulating the Lindbladian dynamics $\frac{d}{dt} \rho(t) = \mathcal{L}\rho(t)$ as $t \to \infty$, the density matrix approaches the steady state which is approximately thermal, and properties of the thermal state can be probed. 
The chiral condensate was measured for varying temperatures and bare masses of the fermion, and the chiral symmetry phase transition was observed numerically to match exact diagonalization results for an $L = 4$ lattice size. 
Transfer learning allows the transfer of learned states on a size $L$ lattice to a size $2L$ lattice, and by successively transfer learning and retraining, the chiral condensate and corresponding phase transition were investigated on an $L = 32$ lattice. 

In principle, the methods in this paper are extendable to realistic simulations of heavy quarkonia propagating through the QGP. 
Neural network simulations of the Lindblad equation can interface with hydrodynamic simulations of the QGP, taking into account the spatiotemporal variations in the temperature at no additional cost to the tVMC algorithm. 
There are various algorithmic challenges associated with this program. 
For example, the Jordan-Wigner transformation applied in the lattice Schwinger Model to transform fermionic variables to spin-$\frac{1}{2}$ variables requires a choice of path through the $3$d-spatial lattice, and will introduce nonlocalities into the Hamiltonian. 
Unless an alternative approach is used, manifest translational symmetry will be lost. 
If a discretisation is used that respects the symmetries of the spatial lattice, it would be interesting to develop an NDO ansatz that respects these symmetries. 
Moreover, existing studies of the Lindbladian dynamics of heavy quarkonia often match the theory onto Non-Relativistic QCD (NRQCD), which has separate particle and antiparticle number conservation.  
The Schwinger Model, on the other hand, only has net particle number conservation.
This causes the Hilbert space of NRQCD to have a different structure (it is not the tensor product of local Hilbert spaces), and may require a different neural network parametrisation than that used here. 
Overcoming these challenges will allow for a more controlled understanding of the dynamic behaviour of heavy quarkonia in the QGP, and by extension a quantitative understanding of the medium itself.

\acknowledgments

We thank Roland Farrell, Bruno Scheihing Hitschfeld, Krishna Rajagopal, Martin Savage and Marc Illa Subina for valuable comments on the manuscript. 
PES and JL are supported in part by the U.S. Department of Energy, Office of Science, Office of Nuclear Physics, under grant Contract Numbers DE-SC0011090, by Early Career Award DE-SC0021006 and by the Simons Foundation grant 994314 (Simons Collaboration on Confinement and QCD Strings).
PES and DL acknowledge the support of the U.S. Department of Energy, Office of Science, National Quantum Information Science Research Centers, Co-design Center for Quantum Advantage (C2QA) under contract number DE-SC0012704. 
X.Y. is supported by the U.S. Department of Energy, Office of Science, Office of Nuclear Physics, InQubator for Quantum Simulation (IQuS) (https://iqus.uw.edu) under Award Number DOE (NP) Award DE-SC0020970 via the program on Quantum Horizons: QIS Research and Innovation for Nuclear Science.
This work is supported by the U.S.\ National Science Foundation under Cooperative Agreement PHY-2019786 (The NSF AI Institute for Artificial Intelligence and Fundamental Interactions, \url{http://iaifi.org/}).
The authors acknowledge the MIT SuperCloud and Lincoln Laboratory Supercomputing Center~\cite{reuther2018interactive} for providing HPC resources that have contributed to the research results reported within this paper.
Numerical experiments and data analysis used Netket\cite{netket3:2021,netket2:2019}, mpi4jax\cite{mpi4jax:2021}, JAX\cite{jax2018github}, NumPy~\cite{harris2020array}, and SciPy~\cite{2020SciPy-NMeth} were used in this study. Figures were produced using matplotlib~\cite{Hunter:2007}.

\appendix

\section{Quantum Numbers}
\label{app:qn}
\begin{figure}[h]
\centering
\begin{subfigure}{0.4\textwidth}
\vspace{4.75em}
\begin{equation*}
\begin{array}{c}
~~~~~~[Q,H]=[Q,F_l]=0~~~\forall l\\
\begin{array}{rcl}
\Neswarrow ~~&  & ~~~\Searrow\\
\frac{d}{dt}Q=\mathcal{L}^\dagger(Q)=0~~ & \Rightarrow & ~~~[\mathcal{Q},\mathcal{L} ]=0
\end{array}
\end{array}
\end{equation*} 
    \caption{Relationship between discrete conserved charges $Q$ and dynamical symmetries \cite{PhysRevA.89.022118}. $F_l$ are the diagonalized jump operators corresponding to the Lindbladian $\mathcal{L}$, see \Cref{eq:jump}. $\mathcal{Q} = Q \otimes Q$ is the charge super-operator, and $\mathcal{L}^\dagger (Q)$ notates the Heisenberg evolution of $Q$. }
    \label{fig:CP_comm}
\end{subfigure}
\hspace{2em}
\begin{subfigure}{0.4\textwidth}
\includegraphics[width=1.3\textwidth]{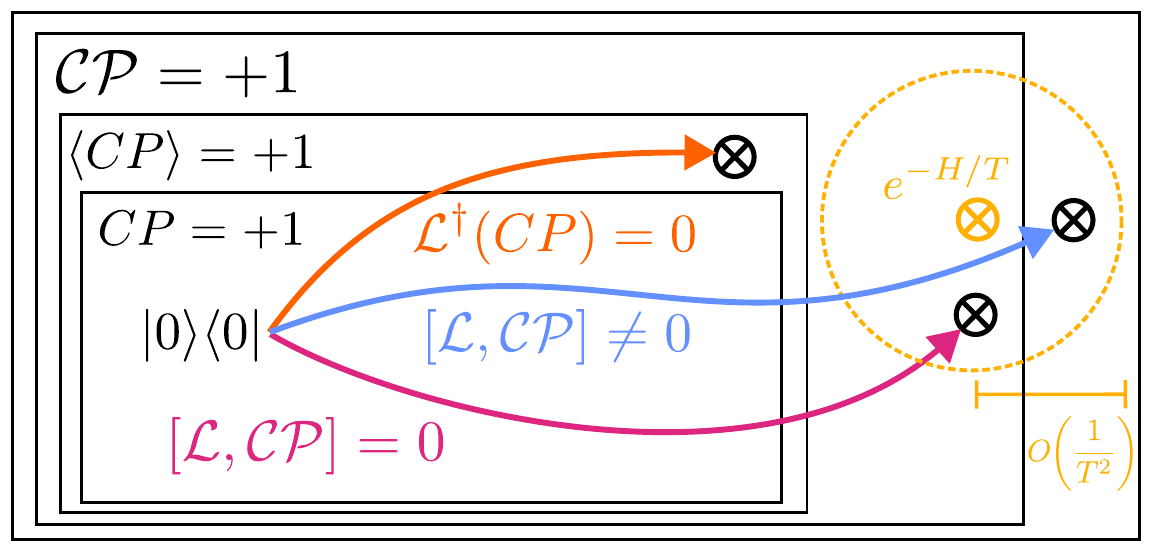}
    \caption{Different evolutions and steady states depending on the $CP$-nature of the coupling matrix. The pictured initial state is the free ground state. Orange depicts strong $CP$ conservation, maroon depicts weak $CP$ conservation and blue depicts no $CP$ conservation.}
    \label{fig:CP_dyn}
\end{subfigure}
\hfill
\vspace{0em}
\begin{subfigure}{1\textwidth}
\begin{center}
\includegraphics[width=0.7\textwidth]{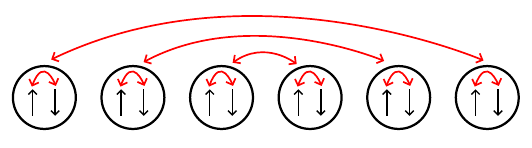}
\end{center}
\vspace{-1.5em}

\caption{Schematic of the ${CP}$ operator, which is a combination of a bond-reflection $S$ and swapping $\uparrow,\downarrow$ on each staggered site, where $\uparrow,\downarrow$ represent the spin-$\frac{1}{2}$ basis of $\sigma^z$. Note that there are no individual $C,P$ symmetries - in particular $[H,{S}] \neq 0$, and $[{H},\prod \sigma_x] \neq 0$.  }
\label{fig:CP_op}
\end{subfigure}
\caption{Details of the CP operator for the discretised Schwinger Model.}
\label{fig:CPfig}
\end{figure}

There are two charge operators that commute with the Schwinger Model Hamiltonian $H$ presented in \Cref{eq:schwinger_ham_1,eq:schwinger_ham_2}. 
The first is the net particle number operator $N = \sum_{i=0}^{L-1} \sigma^z_i$ that counts the number of particles minus the number of antiparticles in the state. 
The second is the $CP$-operator (shown schematically in \Cref{fig:CP_op}) $CP = S \cdot  (\prod^{L-1}_{i=0} \sigma^x_i)$ where $S$ is the unitary operator that swaps all sites $i$ with $L - i + 1$. 
By simultaneously diagonalizing, the total Hilbert space of dimension $2^{L}$ is partitioned into fixed $N$ and $CP$ sectors, where both the free and interacting vacuum states live in the $N = 0, CP = +1$ sector.

For Hamiltonian evolution of pure states, there is only one notion of charge conservation: $([Q,H] = 0) \iff (\frac{d}{dt} Q = 0)$. As shown in  \Cref{fig:CP_comm}, the corresponding statement for super-operators in Lindbladian evolution breaks down, and there are now two different notions of charge conservation. A charge operator $Q$ is `strongly-conserved' if $\frac{d}{dt} Q = 0$, or `weakly-conserved' if $[\mathcal{Q}, \mathcal{L}] = 0$, where $\mathcal{Q}$ is the corresponding superoperator of $Q$ (strong-conservation implies weak-conservation, but not the other way around). \Cref{fig:CP_dyn} shows the different behaviours under Lindbladian evolution produced in each case. The inner-most box represents $CP = +1$ pure states which are used in \Cref{sec:res} as initial states in the studies of Lindbladian evolution, for example the free vacuum state $|0\rangle \langle 0 |$ or the string product state $|e^+(x) e^-(\frac{L}{2} - x) \rangle \langle e^+(x) e^-(\frac{L}{2} - x)| $. The center box $\langle CP \rangle = +1$ represents density matrices $\rho$ where $\mathrm{Tr}(CP \cdot \rho) = 1$, and the outer box $\mathcal{CP} = +1$ denotes density matrices that satisfy $\mathcal{CP} \rho = CP \cdot \rho \cdot CP = \rho$. 

The different conservation properties lead to different long-term behaviour under Lindbladian dynamics, in particular the steady state of the Lindbladian lives in different sectors, as shown in \Cref{fig:CP_dyn}. Strong $CP$ conservation causes $\langle CP \rangle$ to be conserved during Lindbladian evolution. For such a Lindbladian there are two steady states, one with $\langle CP \rangle = +1$ and the other with $\langle CP \rangle= -1$ which can be thought of as restrictions of the thermal state into the two $CP$ sectors. Weak $CP$ conservation preserves the $\mathcal{CP}$-sector that the initial state lives in. So long as the initial state is in the $\mathcal{CP} = +1$ sector, Lindbladian evolution is guaranteed to approach the thermal state as the thermal state also lives in the $\mathcal{CP} = +1$ sector. Finally, in the case that there is no $CP$ conservation, the unique steady state will still be approximately thermal. 

For the choice of coupling operators $O_i = (\overline{\psi} \psi)_i$, the number operator $N$ is always strongly-conserved, regardless of the choice of coupling matrix $D_{ij}$. In particular this means that the Lindbladian dynamics has no direct access to chemical potential effects, as the different fixed-$N$ sectors are decoupled. It is still possible to indirectly probe chemical potential effects by simulating the different fixed-$N$ sectors and appropriately reweighting. The $CP$ super-operator $\mathcal{CP} = CP \otimes CP$ splits the space of density matrices into $\mathcal{CP}$-even and $\mathcal{CP}$-odd sectors. Note that regardless of the $CP$-nature of a pure state, they are all $\mathcal{CP}$-even. As long as the $CP$-operator of the subsystem can be extended to a unitary $CP_\mathrm{tot}$ symmetry of the entire Hilbert space (assuming the medium is $CP$-symmetric), then there is weak $CP$-conservation up to corrections that vanish in the Quantum Brownian Motion limit:

\begin{equation}
\begin{split}
[CP_\mathrm{tot},H_\mathrm{tot}] = 0 &\implies CP_\mathrm{tot} e^{i H_\mathrm{tot}t } |\psi\rangle =  e^{i H_\mathrm{tot}t } CP_\mathrm{tot} |\psi\rangle \\
&\implies \mathcal{CP} e^{\mathcal{L} t} \ \mathrm{Tr}_\mathrm{env}(|\psi\rangle \langle \psi | ) \approx  e^{\mathcal{L} t} \mathcal{CP} \ \mathrm{Tr}_\mathrm{env}(|\psi\rangle \langle \psi | ) \implies [\mathcal{CP},\mathcal{L}] \approx 0.
\end{split}
\end{equation}

This requirement rules out certain choices of coupling matrix $D_{ij}$ that are not weakly-CP conserving. For example, a previously considered ansatz for the coupling matrix is an open boundary condition Gaussian $D_{ij} = D_0 \ \mathrm{exp}(-\frac{1}{2}\frac{|i - j|^2}{w^2})$ for some width $w > 0$ \cite{Lee:2023urk}, however this $D$-matrix violates the weak CP-preservation condition. The physical implication is that the time-evolved electric fields no longer satisfy the symmetry $\langle {E}_i \rangle(t) = \langle E_{L - i} \rangle(t)$. A modified ansatz that satisfies weak CP-preservation is the periodic Gaussian coupling: 
\begin{equation}
D^g_{ij}(w) := D_0 \sum_{n = -\infty}^{\infty} \frac{1}{w \sqrt{2 \pi} } \ e^{-\frac{1}{2} \left(\frac{nL + i - j}{w} \right)^2 } =  \frac{D_0}{L} \vartheta_3\left((i - j) \cdot \frac{\pi}{L},\  e^{ - \frac{2 \pi^2 w^2}{L^2}} \right) 
\end{equation}
which can be expressed in terms of the Jacobi-theta function $\vartheta_3$.
The parameter $D_0$ controls the overall coupling strength, and $w$ is the width of the Gaussian. In the general case with $w$ a finite number, the $D$ matrix can be diagonalized numerically on the open-boundary condition lattice to obtain a list of $L$ jump-operators: 
\begin{equation}
D = U^\dagger \cdot \mathrm{diag}(\lambda_1,\cdots,\lambda_n)\cdot  U, \quad F_i = \sqrt{\lambda_i} U_{ij} \tilde{O}_j, 
\end{equation}
\begin{equation}\label{eq:jump}
\mathcal{L} \rho(t) = - i [H,\rho(t)] + a^2 \sum_i \left( F_i \rho(t) F_i^\dagger - \frac{1}{2} \{ F_i^\dagger F_i, \rho(t) \} \right).  
\end{equation}

The couplings studied in \Cref{sec:real} are limits of the periodic Gaussian coupling, with the delta coupling being the limit as $w \to 0$ and the constant coupling being the limit as $w \to \infty$:
\begin{equation}
\lim_{w \to 0} w \sqrt{2 \pi} \cdot D^g_{ij}(w) =  {D_0 \delta_{ij}}, \quad \lim_{w \to \infty} L\cdot  D^g_{ij}(w) = D_0 \mathbf{1}_{ij}. 
\end{equation}

\bibliographystyle{jhep}
\bibliography{main.bib}
\end{document}